\documentclass[prl,twocolumn,showpacs,preprintnumbers,amsmath,amssymb]{revtex4}
\usepackage{graphicx}
\usepackage{dcolumn}
\usepackage{bm}
\begin{document}
\title{Effective field theory: A complete relativistic nuclear model}
\author{P. Arumugam} 
\author{B.K. Sharma} 
\author{P.K. Sahu} 
\author{S.K. Patra} 
\affiliation{Institute of Physics, Sachivalaya Marg, Bhubaneswar -
751 005, India.}
\date{August 19 2003}

\begin{abstract}
We analyzed the results for finite nuclei and infinite nuclear and
neutron matter using the standard $\sigma-\omega$ model and with
the effective field theory. For the first time, we have shown here
quantitatively that the inclusion of self-interaction of the
vector mesons and the cross-interaction of all the mesons taken in
the theory explain naturally the experimentally observed softness
of equation of state without loosing the advantages of standard
$\sigma-\omega$ model for finite nuclei.  Recent experimental
observations support our findings and allow us to conclude that
without self- and cross-interactions the relativistic mean field
theory is incomplete.
\end{abstract}

 \maketitle

In the quest for an unified model describing both finite nuclei
and nuclear matter, the quantum hadrodynamics (QHD) has been a
successful tool for the past few decades.  QHD is the field theory
of the nuclear many-body problem using hadron degrees of freedom.
Models based on relativistic QHD takes care \textit{ab initio} of
many natural phenomena which are practically absent or have to be
included in an \textit{ad hoc} manner in the non-relativistic
formalism.  One of the first successful models based on QHD with
relativistic mean field (RMF) was constructed by Walecka
\cite{Wa74} with vector and scalar meson fields. Later on, to get
reasonable incompressibility and to get good results for finite
nuclei, cubic and quartic nonlinearities of the $\sigma$ meson
(standard nonlinear $\sigma -\omega$ model) were added
\cite{Bo77}. These models were proposed to be renormalizable and
that constraint limited the scalar interactions to a quartic
polynomial and disallowed the scalar-vector cross interactions and
vector-vector self-interactions.  However, the coupling constants
are not assigned with their bare (experimental) values but with
some effective value to have proper results for finite nuclei.
Hence the renormalizability of the Lagrangian gets compromised by
the use of effective coupling constants.

Inspired by effective field theory (EFT), Furnstahl, Serot and
Tang \cite{Fu96} abandoned the idea of renormalizability and
extended the RMF theory by allowing other nonlinear scalar-vector
and vector-vector interactions in addition to tensor couplings
\cite{Fu96,Se97,Mu96,Fu96b,Fu00}. The EFT contains all the
non-renormalizable couplings consistent with the underlying
symmetries of QCD.  The effective Lagrangian is obtained by
employing suitable expansion scheme to truncate the infinite
number of terms.  In such scheme the ratios $\Phi/M$, $W/M$,
$|\mbox{\boldmath $\nabla$}\Phi|/M^2$ and $|\mbox{\boldmath
$\nabla$} W|/M^2$ are the useful expansion parameters
\cite{Fu96,Se97,Mu96,Es99,estal1} where $\Phi$ and $W$ are scalar
and vector meson fields respectively and $M$ is the nucleon mass.
With the help of the concept of naturalness (i.e., all coupling
constants are of the order of unity when written in appropriate
dimensionless form), it is then possible to compute the
contributions of the different terms in the expansion and to
truncate the effective Lagrangian at a given level of accuracy
\cite{Fu96,Se97,Fu96b,Fu00}. None of the couplings should be
arbitrarily dropped out to the given order without a symmetry
argument.  References \cite{Fu96b,Fu00,Es99,estal1} have shown
that it suffices to go to fourth order in the expansion. At this
level one recovers the standard nonlinear $\sigma-\omega$ model
plus a few additional couplings, with thirteen free parameters in
all. These parameters have been fitted (parameter sets G1 and G2)
to reproduce some observables of magic nuclei \cite{Fu96}. The
fits display naturalness, and the results are not dominated by the
last terms retained. This evidence confirms the utility of the EFT
concepts and justifies the truncation of the effective Lagrangian
at the first lower orders.

Recent applications of the models based on EFT include studies of
pion-nucleus scattering \cite{Cl98} and of the nuclear spin-orbit
force \cite{Fu98}, as well as calculations of asymmetric nuclear
matter at finite temperature with the G1 and G2 sets \cite{Wa00}.
In a previous work \cite{Es99} we have analyzed the impact of each
one of the new couplings introduced in the EFT models on the
nuclear matter saturation properties and on the nuclear surface
properties. In Ref.\ \cite{estal1} we have looked for constraints
on the new parameters by demanding consistency with
Dirac-Brueckner-Hartree-Fock (DBHF) \cite{DDHF} calculations and
the properties of finite nuclei. Using EFT we successfully
explained the properties of the drip-line nuclei as well as the
symmetric and asymmetric infinite nuclear matter including the
neutron star and compared with other theoretical calculations
\cite{BKS}.  Very recently \cite{Sci02}, the flow of matter in
heavy-ion collisions is analyzed to determine the pressures
attained at densities ranging from two to five times the
saturation density of nuclear matter. This experimental
determination of the equation of state (EOS) of dense matter
motivated us to study the applicability of various relativistic
models at extreme conditions.  In this letter we present the
observations based on the results of our EFT calculations, the
standard nonlinear $\sigma-\omega$ model and some of the recent
experiments.

The description of EFT and the field equations for nuclear matter
and finite nuclei can be found in Refs. \cite{Se97,Fu96}. The
field equations were derived \cite{Fu96}  from an energy density
functional containing Dirac baryons and classical scalar and
vector mesons. According to Refs. \cite{Se97,Fu96} the energy
density for finite nuclei can be written as
\begin{widetext}
\begin{eqnarray}
{\cal E}(\textbf{r}) & = &  \sum_\alpha \varphi_\alpha^\dagger
\Bigg\{ -i \mbox{\boldmath$\alpha$} \!\cdot\!
\mbox{\boldmath$\nabla$} + \beta (M - \Phi) + W +
\frac{1}{2}\tau_3 R + \frac{1+\tau_3}{2} A - \frac{i}{2M} \beta
\mbox{\boldmath$\alpha$}\!\cdot\! \left( f_v
\mbox{\boldmath$\nabla$} W + \frac{1}{2}f_\rho\tau_3
\mbox{\boldmath$\nabla$} R + \lambda \mbox{\boldmath$\nabla$} A
\right)
\nonumber \\
& & + \frac{1}{2M^2}\left (\beta_s + \beta_v \tau_3 \right )
\Delta A \Bigg\} \varphi_\alpha \null + \left ( \frac{1}{2} +
\frac{\kappa_3}{3!}\frac{\Phi}{M} +
\frac{\kappa_4}{4!}\frac{\Phi^2}{M^2}\right )
\frac{m_{s}^2}{g_{s}^2} \Phi^2  - \frac{\zeta_0}{4!} \frac{1}{
g_{v}^2 } W^4
\nonumber \\
& & \null + \frac{1}{2g_{s}^2}\left( 1 +
\alpha_1\frac{\Phi}{M}\right) \left( \mbox{\boldmath
$\nabla$}\Phi\right)^2 - \frac{1}{2g_{v}^2}\left( 1
+\alpha_2\frac{\Phi}{M}\right) \left( \mbox{\boldmath $\nabla$} W
\right)^2 \null - \frac{1}{2}\left(1 + \eta_1 \frac{\Phi}{M} +
\frac{\eta_2}{2} \frac{\Phi^2 }{M^2} \right)
\frac{{m_{v}}^2}{{g_{v}}^2} W^2
\nonumber \\
& &  - \frac{1}{2g_\rho^2} \left( \mbox{\boldmath $\nabla$}
R\right)^2 - \frac{1}{2} \left( 1 + \eta_\rho \frac{\Phi}{M}
\right) \frac{m_\rho^2}{g_\rho^2} R^2 \null - \frac{1}{2e^2}\left(
\mbox{\boldmath $\nabla$} A\right)^2 + \frac{1}{3g_\gamma g_{v}}A
\Delta W + \frac{1}{g_\gamma g_\rho}A \Delta R , \label{eqFN1}
\end{eqnarray}
\end{widetext}
where the index $\alpha$ runs over all occupied states
$\varphi_\alpha (\textbf{r})$ of the positive energy spectrum,
$\Phi \equiv g_{s} \phi_0(\textbf{r})$, $ W \equiv g_{v}
V_0(\textbf{r})$, $R \equiv g_{\rho}b_0(\textbf{r})$, $A \equiv e
A_0(\textbf{r})$. $g_s,\ g_v,\ g_{\rho}$ and $e$ are the coupling
constants corresponding to the fields $\phi_0(\textbf{r})$, $
V_0(\textbf{r})$, $b_0(\textbf{r})$ and $A_0(\textbf{r})$
respectively and $\kappa_3$, $\kappa_4$, $\eta_1$, $\eta_2$,
$\zeta_0$, $f_v$ and $f_{\rho}$ are non-linear coupling constants.

The terms with $g_\gamma$, $\lambda$, $\beta_{s}$ and $\beta_{v}$
take care of effects related with the electromagnetic structure of
the pion and the nucleon (see Ref.\ \cite{Fu96}). Specifically,
the constant $g_\gamma$ concerns the coupling of the photon to the
pions and the nucleons through the exchange of neutral vector
mesons. The experimental value is $g_\gamma^2/4\pi = 2.0$. The
constant $\lambda$ is needed to reproduce the magnetic moments of
the nucleons. It is defined by
\begin{eqnarray}
\lambda = \frac{1}{2} \lambda_{p} (1 + \tau_3) + \frac{1}{2}
\lambda_{n} (1 - \tau_3) , \label{eqFN2}
\end{eqnarray}
with $\lambda_{p} = 1.793$ and $\lambda_{n}=-1.913$ the anomalous
magnetic moments of the proton and the neutron, respectively. The
terms with $\beta_{s}$ and $\beta_{v}$ contribute to the charge
radii of the nucleon \cite{Fu96}.

Variation of the energy density (\ref{eqFN1}) with respect to
$\varphi^\dagger_\alpha$ and the meson fields gives the Dirac
equation fulfilled by the nucleons and the meson field equations
\cite{estal1}.  The Dirac equation corresponding to the energy
density (\ref{eqFN1}) and the mean field equations for $\Phi$,
$W$, $R$ and $A$ can be found in Ref. \cite{estal1,BKS}. The meson
fields can also be interpreted as Kohn--Sham potentials
\cite{Ko65} in the relativistic case \cite{Sp92} and in this sense
they include effects beyond the Hartree approach like three-body
and many-body interactions through the nonlinear couplings
\cite{Fu96,Se97}.

For infinite nuclear matter all of the gradients of the fields in
the energy density and field equations vanish. Due to the fact
that the solution of symmetric and asymmetric nuclear matter in
mean field depends on the ratios $g_s^2/m_s^2$ and $g_v^2/m_v^2$
\cite{Se86}, we have seven unknown parameters. By imposing the
values of the saturation density, total energy, incompressibility
modulus and effective mass, we still have three free parameters
(the value of $g_\rho^2/m_\rho^2$ is fixed from the bulk symmetry
energy coefficient $J$).

The baryon, scalar, isovector, proton and tensor densities are
same as for finite nuclei in the nuclear matter limit, i.e.,
$\rho=\frac{\gamma}{(2\pi)^3}\int_0^{k_f}{d^3}k=\frac{\gamma}{6{\pi^2}}{k_f^3}$
and
$\rho_s=\frac{\gamma}{(2\pi)^3}\int_0^{k_f}{d^3}k\frac{M^{*}}{\sqrt{({k^2}+{M^*{^2}})}}$,
and so on (here $k_f$ is the Fermi momentum and $k$ is the
momentum at any density).  The expressions for pressure and energy
density are
\begin{widetext}
\begin{eqnarray}
P & = &\frac{\gamma}{3(2\pi)^3}\int
d^3k\frac{k^2}{E^*(k)}+\frac{1}{4!}\zeta_0g_w^2V_0^4
+\frac{1}{2}\Bigg(1+\eta_1\frac{g_{\sigma}\sigma}{M}+\frac{\eta_2}{2}
\frac{g_{\sigma}^2\sigma^2}{M^2}\Bigg)m_{\omega}^2V_{0}^{2}
\nonumber \\
& & \null -m_{\sigma}^2\sigma^2\Bigg(\frac{1}{2}+\frac{\kappa_3
g_{\sigma}\sigma}{3! M} +\frac{\kappa_4 g_{\sigma}^2\sigma^2}{4!
M^2}\Bigg)+ \frac{1}{2}
\Bigg(1+\eta_{\rho}\frac{g_{\sigma}\sigma}{M}\Bigg)m_{\rho}^2b_{0}^{2}
\;, \label{eqFN25}
\end{eqnarray}

\begin{eqnarray}
\epsilon & = &\frac{\gamma}{(2\pi)^3}\int
d^3kE^*(k)-\frac{1}{4!}\zeta_0g_w^2V_0^4
-\frac{1}{2}\Bigg(1+\eta_1\frac{g_{\sigma}\sigma}{M}+\frac{\eta_2}{2}
\frac{g_{\sigma}^2\sigma^2}{M^2}\Bigg)m_{\omega}^2V_{0}^{2} +
\frac12 g_{\rho}b_0(\rho_p-\rho_n)
\nonumber \\
& & \null +m_{\sigma}^2\sigma^2\Bigg(\frac{1}{2}+\frac{\kappa_3
g_{\sigma}\sigma}{3! M} +\frac{\kappa_4 g_{\sigma}^2\sigma^2}{4!
M^2}\Bigg)+ \frac{1}{2}
\Bigg(1+\eta_{\rho}\frac{g_{\sigma}\sigma}{M}\Bigg)m_{\rho}^2b_{0}^{2}
 +g_{\omega}V_{0}(\rho_p+\rho_n) \;.\label{eqFN26}
\end{eqnarray}
\end{widetext}
Here $\gamma$=2 for pure neutron matter and $\gamma$=4 for
symmetric nuclear matter. The asymmetry of nuclear matter is
defined by the parameter $\alpha$. For the symmetric matter,
$\alpha$=0 and for the neutron matter, $\alpha$=1.

While examining the effect of nonlinear coupling in nuclear matter
\cite{estal1} using various nuclear force parameters, the NL3
parameter set \cite{Lal97} (considered as a representative of the
standard nonlinear $\sigma-\omega$ parametrization) has been found
to fail in following the DBHF results even at slightly higher
densities. It is well known that the DBHF theory in relativistic
framework explains well the nuclear matter at higher densities
($\sim 2\rho_0$) \cite{Broc90,Sug94}.  In contrast to the standard
$\sigma-\omega$ model, the EFT calculations at high density
regimes yield results in accordance with DBHF. This scenario is
depicted in Fig.~1, where we present the results of calculations
with TM1 parameter set \cite{Sug94} also. In the calculation with
TM1, only a quartic vector self-interaction term is included apart
from the terms in NL3. This inclusion was done arbitrarily without
considering the underlying QCD symmetries or the naturalness.
However, with this self-interaction term, the TM1 give better
results at higher densities.  This demonstrates the importance of
self-interactions at higher densities and exposes the inadequacy
of the standard nonlinear $\sigma-\omega$ model. This argument is
further supported by Fig. 2 in which the variation of binding
energy per particle ($E/A$) is plotted as a function of
$\rho/\rho_0$. The NL3 parameter set gives a much too stiff EOS
whereas the other parameter sets give a softer EOS which is
consistent with the observed neutron star masses \cite{Mu96} and
radii (See Table I) and measurements of kaon production in
heavy-ion collisions \cite{PRL01}.

\begin{figure}
\includegraphics[width=0.75\columnwidth, clip=true]{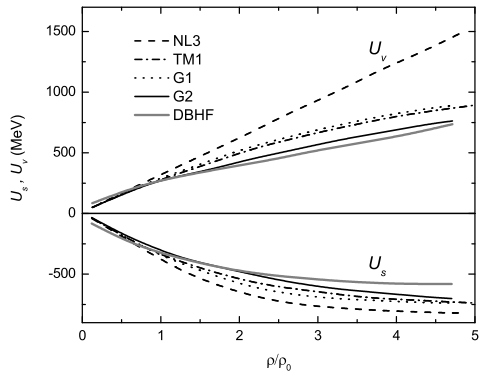}
\caption{\label{fig1} Scalar ($U_s$) and vector ($U_v$) potentials
against the ratio of nuclear matter density with its saturation
value ($\rho_0$), using the parametrizations NL3 \cite{estal1},
TM1 \cite{Sug94}, G1, and G2 \cite{Fu96} and in a DBHF calculation
\cite{DDHF}.}
\end{figure}

\begin{figure}
\includegraphics[width=0.75\columnwidth, clip=true]{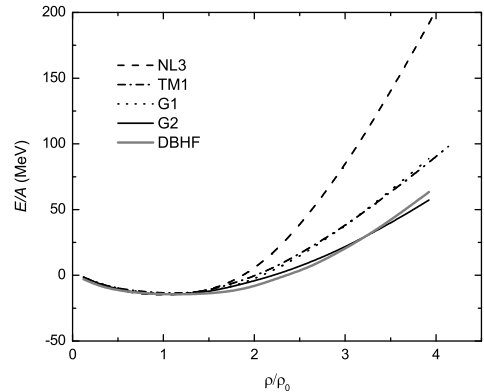}
\caption{\label{fig2} EOS for the same cases as in Fig. 1.}
\end{figure}

\begin{table}
\caption{\textit{Upper panel}: The surface energy coefficient
$E_s$ and surface thickness $t$ (in fm).  \textit{Middle panel}:
Energy per nucleon $E/A$ (in MeV), charge radius $r_{\rm ch}$ (in
fm) and spin-orbit splittings $\Delta E_{\rm SO}$ (in MeV) of the
least-bound nucleons. \textit{Lower panel}: the neutron star
radius $R$ (in km) and the mass ratio $M/M_\odot$.}
\begin{tabular}{llrrrrr}
\hline \hline
        &       &    TM1    &    NL3    &    G1     &    G2     &    Exp. \\
\hline
    &    $E_s$  &   18.51   &   18.36   &   18.06   &   17.8    &    16.5--21.0\\
    &    $t$    &   1.91    &   1.99    &   1.98    &   2.08    &    2.2--2.5\\
\hline $^{16}$O     &    $E/A$      &    $-$8.15    &    $-$8.08    &    $-$7.97    &    $-$7.97    &     $-$7.98\\
    &    $r_{\rm ch}$   &   2.66    &   2.73    &   2.72    &   2.72    &    2.73 \\
    &    $\Delta E_{\rm SO}$ (n,$1p$)   &   5.6 &   6.4 &   6   &   5.9 &    6.2 \\
    &    \hspace{.87cm} (p,$1p$)    &   5.6 &   6.3 &   5.9 &   5.9 &    6.3 \\
$^{48}$Ca   &    $E/A$      &    $-$8.65    &    $-$8.64    &    $-$8.67    &    $-$8.68    &     $-$8.67 \\
    &    $r_{\rm ch}$   &   3.46    &   3.48    &   3.44    &   3.44    &    3.47 \\
    &    $\Delta E_{\rm SO}$ (n,$1d$)   &   5   &   6.1 &   5.8 &   5.6 &    3.6\\
    &    \hspace{.87cm} (p,$1d$)    &   5.2 &   6.3 &   6.2 &   6   &    4.3\\
$^{90}$Zr   &    $E/A$      &    $-$8.71    &    $-$8.69    &    $-$8.71    &    $-$8.68    &     $-$8.71\\
    &    $r_{\rm ch}$   &   4.27    &   4.28    &   4.28    &   4.28    &    4.26 \\
    &    $\Delta E_{\rm SO}$ (n,$2p$)   &   1.4 &   1.6 &   1.8 &   1.8 &    0.5 \\
$^{208}$Pb  &    $E/A$      &    $-$7.87    &    $-$7.87    &    $-$7.87    &    $-$7.86    &     $-$7.87\\
    &    $r_{\rm ch}$   &   5.54    &   5.52    &   5.5 &   5.5 &    5.50\\
    &    $\Delta E_{\rm SO}$ (n,$3p$)   &   0.7 &   0.8 &   0.9 &   0.9 &    0.9\\
    &    \hspace{.87cm} (p,$2d$)    &   1.4 &   1.6 &   1.8 &   1.8 &    1.3\\
\hline
    &   $R$ &   13.91   &   15.47   &   13.93   &   10.04   &    10.0--12.0\\
    &   $M/M_\odot$ &   2.8 &   2.78    &   2.16    &   2.04    &    1.5--2.5\\
\hline \hline
\end{tabular}
\end{table}

\begin{figure}
\includegraphics[width=0.75\columnwidth, clip=true]{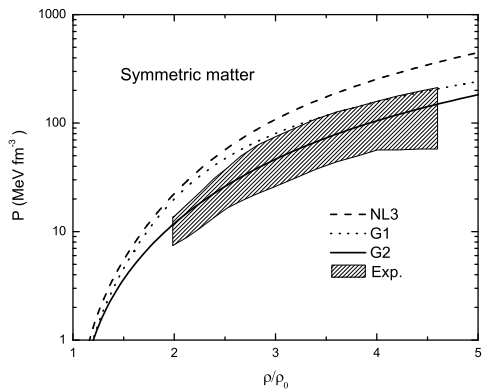}
\caption{\label{fig3} Zero temperature EOS for symmetric nuclear
matter.  The shaded region corresponds to experimental data
\cite{Sci02}.  The EOS from calculations using NL3, G1 and G2
parameter sets are represented by dashed, dotted and solid lines
respectively.}
\end{figure}

\begin{figure}
\includegraphics[width=0.75\columnwidth, clip=true]{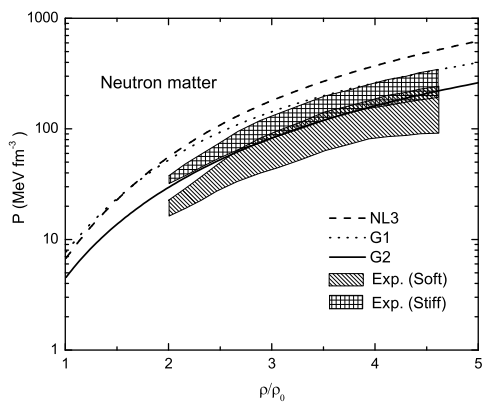}
\caption{\label{fig4} Zero temperature EOS for neutron matter. The
upper and lower shaded regions correspond to experimental data
\cite{Sci02} with strong and weak density dependences,
respectively. Other cases are same as in Fig. 3.}
\end{figure}

The recent experimental observations \cite{Sci02,PRL01} rule out
any strongly repulsive nuclear EOS and has confirmed the
predictions made above. The zero-temperature EOS for symmetric
nuclear matter derived experimentally is shown in Fig. 3 along
with the results obtained from different calculations.   In Fig. 3
we can see that the calculations based on NL3 deviate drastically
from the experimental observation and the EFT calculations with G1
upto some extent and G2 to an excellent extent, agree with the
experiment. A similar situation prevails in the EOS of neutron
matter and can be seen in Fig. 4. From the figures it is very
clear that NL3 calculations are not suitable for nuclear matter as
the formalism is incomplete without the self- and
cross-interaction terms. On the other hand the EFT calculations
with G1 and G2 parameter sets explain the situation in nuclear
matter naturally without any forced changes in parameters or in
formalism and the fit with the experiment is outstanding.  The
experimental data are explained reasonably good only by the
calculations of Akmal \textit{et al} \cite{Akmal} which employs
the Argonne $v_{18}$ interaction.  Such an interaction is not
applied successfully in the case of finite nuclei.  The EFT
calculations are proved to give very good results for finite
nuclear properties \cite{estal1} as well as for infinite nuclear
matter including neutron star \cite{BKS}. In Table I we present
some of the sample results of EFT calculations. More results for
finite nuclei and nuclear matter can be found in Ref.
\cite{estal1,BKS}. In choosing between the parameter sets G1 and
G2 for further calculations we prefer G2. It is worth to note that
G2 presents positive values of $\Phi^4$ coupling constant
($\kappa_4$), as opposed to G1 and to many of the most successful
RMF parametrizations, such as NL3. Actually the negative value of
$\kappa_4$ is not acceptable because the energy spectrum then has
no lower bound \cite{Baym60}. However such negative value is
necessary in the standard $\sigma-\omega$ model to get the results
closer to the experimental values.  On the other hand to have
positive value for $\kappa_4$ it is not necessary to make two
parameter sets as was done in Ref. \cite{Sug94}.

In conclusion, with the new experimental values for EOS, the
predictions of EFT are proved to be true.  With the inclusion of
self- and cross-interactions and without forcing any change of
parameters or modifying the formalism the EFT calculations with G2
parameter set explain finite nuclei and infinite nuclear matter in
a unified way with commendable level of accuracy in both the
cases.  Any Lagrangian without all types of self- and
cross-interactions is incomplete and at present EFT can be
considered as a complete unified theory for finite nuclei as well
as for infinite nuclear matter.

\end{document}